\documentclass[aps,prx,twocolumn,showpacs,floatfix,superscriptaddress]{revtex4}

\usepackage{times}
\usepackage{physics}
\usepackage{graphicx}
\usepackage{hyperref}
\hypersetup{colorlinks=true,linkcolor=blue,citecolor=blue}
\usepackage{amsmath}
\usepackage{wrapfig}
\usepackage{siunitx}
\bibliography{Sci,books,arXiv}
\usepackage{float}
\begin{document}

\title{Photoinduced Rashba spin to charge conversion via interfacial unoccupied state}

\begin{abstract}
At interfaces with inversion symmetry breaking, Rashba effect couples the motion of electrons to their spin; as a result, spin – charge interconversion mechanism can occur. These interconversion mechanisms commonly exploit Rashba spin splitting at the Fermi level by spin pumping or spin torque ferromagnetic resonance. Here, we report evidence of significant photoinduced spin to charge conversion via Rashba spin splitting in an unoccupied state above the Fermi level at the Cu(111)/$\alpha$-Bi$_{2}$O$_{3}$ interface. We predict an average Rashba coefficient of $1.72\times 10^{-10}eV.m$ at 1.98 eV above the Fermi level, by fully relativistic first – principles analysis of the interfacial electronic structure with spin orbit interaction. We find agreement with our observation of helicity dependent photoinduced spin to charge conversion excited at 1.96 eV at room temperature, with spin current generation of $J_{s}=10^{6}A/m^{2}$. The present letter shows evidence of efficient spin – charge conversion exploiting Rashba spin splitting at excited states, harvesting light energy without magnetic materials or external magnetic fields.  
\end{abstract}
\date{\today}

\author{Jorge Puebla}
\email{jorgeluis.pueblanunez@riken.jp}
\affiliation{CEMS, RIKEN, Saitama, 351-0198, Japan}
\author{Florent Auvray}
\affiliation{Institute for Solid State Physics, University of Tokyo, 5-1-5 Kashiwanoha, Kashiwa, Chiba, 277-8581, Japan}
\author{Naoya  Yamaguchi}
\affiliation{Division of Mathematical and Physical Sciences, Graduate School of Natural Science and Technology, Kanazawa University, Kanazawa, Ishikawa, 920-1192, Japan}
\author{Mingran Xu}
\affiliation{Institute for Solid State Physics, University of Tokyo, 5-1-5 Kashiwanoha, Kashiwa, Chiba, 277-8581, Japan}
\author{Satria  Zulkarnaen  Bisri} 
\affiliation{CEMS, RIKEN, Saitama, 351-0198, Japan}
\author{Yoshihiro Iwasa} 
\affiliation{CEMS, RIKEN, Saitama, 351-0198, Japan}
\affiliation{Quantum Phase Electronic Center (QPEC) and Department of Applied Physics, University of Tokyo, Bunkyo-ku, Tokyo, 113-8656, Japan}
\author{Fumiyuki Ishii} 
\affiliation{Faculty of Mathematics and Physics, Institute of Science and Engineering, Kanazawa University, Kanazawa, Ishikawa, 920-1192, Japan}
\affiliation{Nanomaterials Research Institute, Kanazawa University, Kakuma-machi, Kanazawa, 920-1192, Japan}
\author{Yoshichika Otani}
\email{yotani@issp.u-tokyo.ac.jp}
\affiliation{CEMS, RIKEN, Saitama, 351-0198, Japan}
\affiliation{Institute for Solid State Physics, University of Tokyo, 5-1-5 Kashiwanoha, Kashiwa, Chiba, 277-8581, Japan}
\maketitle

Rashba effect has provided fertile ground for basic research and innovative device proposals in condensed matter. Particularly attractive is the fact that in crystals lacking spatial inversion symmetry the induced spin orbit field couples to the electron’s magnetic moment. This spin orbit coupling (SOC) allows the conversion of spin current to transverse electrical charge, or vice versa, the conversion of unpolarized electrical current to spin polarization and diffusion as spin current. These mechanisms have been confirmed in a variety of systems lacking spatial inversion symmetry, opening the condensed matter sub-field of spin orbitronics \cite{Otani, Manchon}. Although, the first demonstration of spin – charge interconversion occurred in semiconductor bulk systems, the recent focus has been the lack of spatial inversion symmetry at metal/metal, metal/semiconductor, metal/oxide and oxide/oxide interfaces; as well as surface states in topological insulators \cite{Ando}. Common techniques for exploring the spin – charge interconversion phenomena at interfaces are the spin pumping and spin transfer torque ferromagnetic resonance. These techniques allow studying the conversions at occupied states below the Fermi level. Arguably, the hybrization of states at interfaces of seemingly different material systems lead to a complex modified electronic structure with multiple Rashba SOC crossings below and above the Fermi level, and even topological points. This statement has been tested by evidence showing a significant modulation of SOC as the Fermi level is increased and new states are occupied \cite{Liang, Herranz, Kondou}. However, feasibility of spin to charge conversion via Rashba spin splitting at unoccupied states has been elusive. 

In this letter we show evidence of photoinduced spin to charge conversion via Rashba spin splitting of unoccupied states at the Cu(111)/$\alpha$-Bi$_{2}$O$_{3}$ interface. Recent reports showed the efficient spin - charge interconversion phenomena at the Cu(111)/$\alpha$-Bi$_{2}$O$_{3}$ interface by microwave photon spin pumping \cite{Tsai}, acoustic spin pumping \cite{Mingran} and magneto optical Kerr effect detection of current induced spin polarization \cite{Puebla, Florent}. The origin of the formation of the two dimensional gas (2DEG) with SOC at this interface between polycrystalline layers is an ongoing topic of debate. One leading hypothesis is the formation of 2DEG by interfacial charge transfer facilitated by the presence of a significant concentration of oxygen defects; hypothesis recently proposed as mechanism for the formation of 2DEG at the amorphous / crystalline perovskite oxide interfaces \cite{Li}. We recently reported the properties of the two dimensional electron gas formation in the Cu(111)/$\alpha$-Bi$_{2}$O$_{3}$ interface with spin orbit coupling by spectroscopic ellipsometry \cite{Manuel}. Polycrystalline interfaces have the advantage of reduced interfacial strain and higher carrier concentrations when compared with highly crystalline interfaces. Here, we performed density functional calculations of our Cu(111)/$\alpha$-Bi$_{2}$O$_{3}$ \cite{Suppl}. Figure 1a shows the calculated layer-projected density of states (LDOS) at the Cu(111)/$\alpha$-Bi$_{2}$O$_{3}$ interface (dashed line zone) and its vicinity. At the interface, it is possible to observe a modification of the LDOS at both sides close to the interface, corresponding to Cu and Bi$_{2}$O$_{3}$ hybridization of Cu-O-Bi states due charge transfer. We sketched the calculated charge density of the electronic state of the Cu(111)/$\alpha$-Bi$_{2}$O$_{3}$ interface in figure 1b; where blue, red and purple spheres are Cu, O and Bi atoms, respectively; yellow clouds represent the Cu-O-Bi states.  

\begin{figure}[t!]
\begin{center}
\includegraphics[height=6.0cm, keepaspectratio]{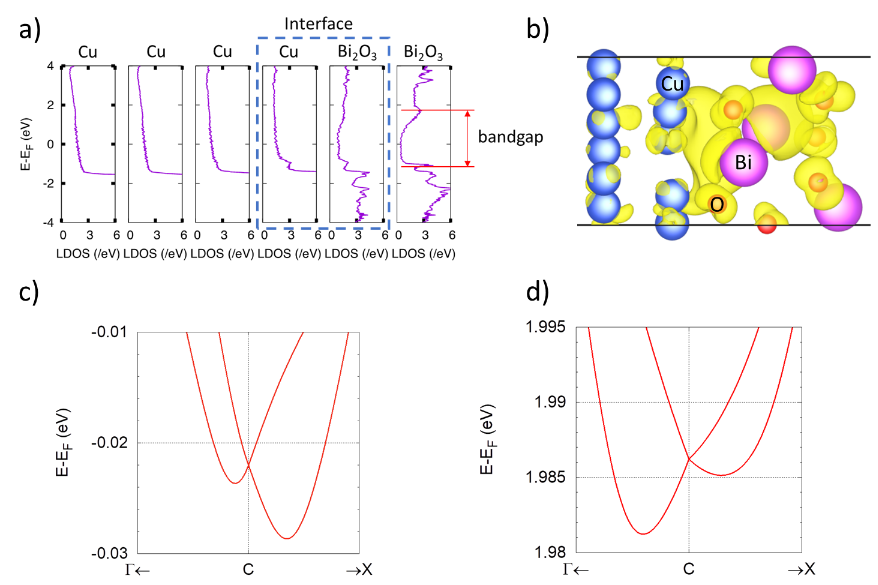}
\end{center}
\caption{First - principles analysis of the Cu(111)/$\alpha$-Bi$_{2}$O$_{3}$ interface. (a) Local density of states at the Cu(111)/$\alpha$-Bi$_{2}$O$_{3}$ interface (dashed line zone) and its vicinity. (b) Schematic representation of the electronic states of the Cu(111)/$\alpha$-Bi$_{2}$O$_{3}$ interface based on density functional theory. Blue spheres are Copper atoms, red Oxygen atoms and purple Bismuth atoms; the yellow shadows show the Cu-O-Bi states. Enlarged views of the band structures around C-point are shown through each path within the range of (c) 0.02 (Bohr)$^{-1}$ and (d) 0.005 (Bohr)$^{-1}$ from C-point along C$\Gamma$ or CX line. The origin in energy set to be the Fermi level and there are special points: $\Gamma$ (0, 0, 0); C (0.5, 0.5, 0); X (0.5, 0, 0).}
\label{fig:1}
\end{figure}

Band structure analysis around the C-point of states below the Fermi level, show a Rashba-like band splitting (see figure 1c). We evaluated Rashba parameters at C point (0.5, 0.5) in the unit of two-dimensional reciprocal lattice vectors. From the spin split energy band dispersion, the Rashba parameter is calculated by using an expression $\alpha_{R}=2$E$_{R}$/k$_{R}$, where E$_{R}$ is the Rashba energy and k$_R$ is the Rashba momentum offset \cite{Ishizaka}. The  averaged Rashba coefficient $\alpha_{R}=0.91$$\times10^{-10}$eV.m accounts for the energy band spin splitting around the Fermi energy as shown in Fig 1 (c), which is same order of magnitude  of  values reported by spin to charge conversion \cite{Mingran, Tsai}. Remarkably, around 1.98 eV above the Fermi level we locate another Rashba splitting of an unoccupied state. The  averaged Rashba coefficient $\alpha_{R}=1.72$$\times10^{-10}$eV.m accounts for the large energy band spin splitting around 1.98 eV above the Fermi energy as shown in Fig 1 (d), almost two times larger than that observed around the Fermi level. The energy of this unoccupied state with Rashba splitting is in close proximity to the well-known interband transition between $d$-states and $s$-states of Cu, with both states participating at the interfacial hybrization.    

We test the photoinduced Rashba spin to charge conversion with excitation energy of 1.96 eV at room temperature. The configuration of our photoinduced spin to charge conversion experiment is sketch in figure 2(a). We generate spin currents by the absorption of angular momentum from light, via the photovoltaic conversion. The angular momentum of light is dictated by its degree of circular polarization or helicity. Notice that, unlike standard heterostructures for photovoltaic devices where photovoltaic collection occurs at bottom and top electrodes, in our device the photovoltaic collection occurs in transverse geometry, following the inverse Edelstein effect (IEE) spin to charge conversion, E$\approx$$\sigma_{s}$$\times$$J_{s}$ \cite{Mingran, Tsai, Rojas}, where $\sigma_{s}$ is the vector of spin polarization and $J_{s}$ is the flow direction of spin current. The interface is formed between a 30 nm thick Cu layer and 20 nm thick Bi$_{2}$O$_{3}$. These thicknesses are selected to suppress interaction of Si/SiO$_{2}$ substrate and the excitation light. The laser beam has an incidence angle $\theta$ and an azimuthal angle $\Psi$. The photon polarization is controlled by a linear polarizer and a quarter wave plate mounted on a rotator.

\begin{figure}[t!]
\begin{center}
\includegraphics[height=10.0cm, keepaspectratio, trim={0 0.3cm 0 0},clip]{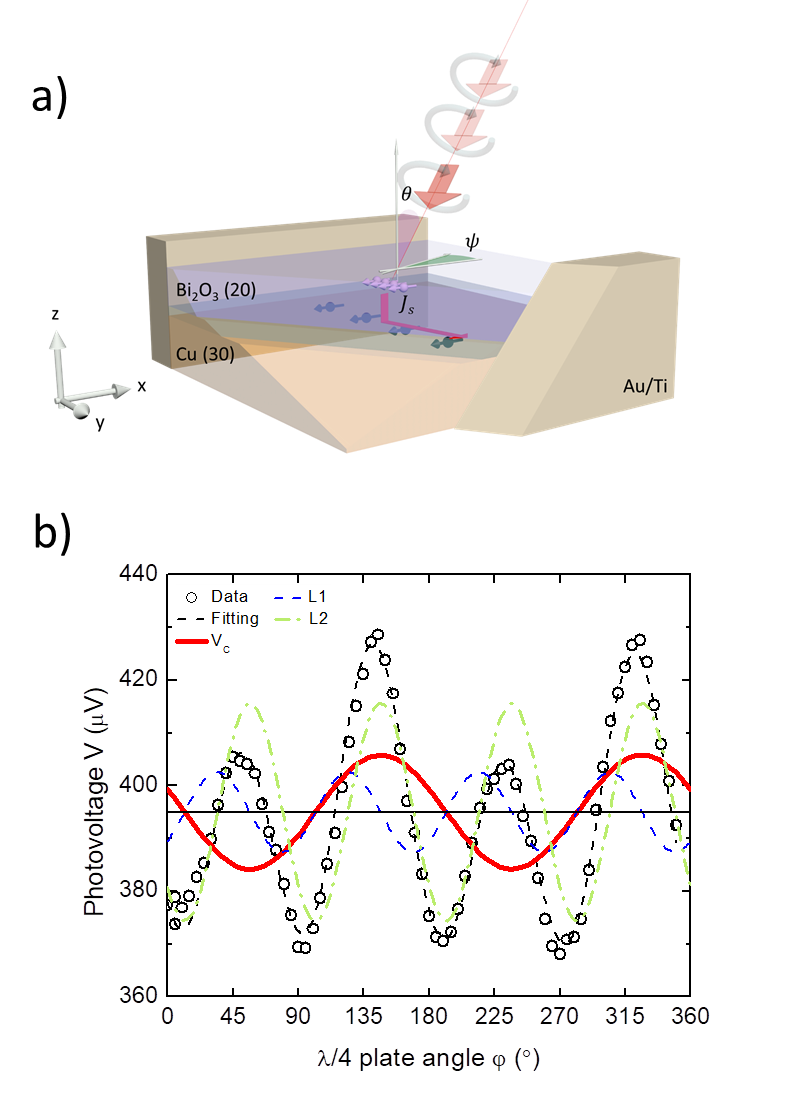}
\end{center}
\caption{(a) Schematic representation of transverse photovoltaic detection following E$\approx$$\sigma_{s}$$\times$$J_{s}$. A laser illuminates the sample at an incidence angle $\theta$ and an azimuthal angle $\Psi$ with polarization $\sigma^{\pm}$. (b) Helicity dependent photoinduced conversion. Photon polarization dependence of transverse photovoltage of Cu(111)/$\alpha$-Bi$_{2}$O$_{3}$ at incidence angle $\theta=70^{\circ}$ and azimuthal angle $\Psi=0^{\circ}$. Open black circles represent the data and dashed black line the fit following eq. (1). Blue dashed line shows the linear polarization contribution ($L_{1}$). Red line shows the circular polarization contribution ($V_{C}$). Green dashed line shows the photovoltage related to Fresnel factors ($L_{2}$).}
\label{fig:2}
\end{figure}

In figure 2(b), we show the helicity dependent photovoltaic measurement obtained with excitation laser energy of 1.96eV. Changing contributions of polarized light due to the rotation of the quarter wave plate ($\varphi$) leads to periodic modulation in photovoltage with a periodicity of 90$^{\circ}$. Photovoltage peaks have different amplitudes, showing periodically two different values. This asymmetry comes from the circularly polarized light modulation analogous to the circular photogalvanic effect \cite{Ganichev}. To better describe the contributions in our modulated signal, we fit the data with the following phenomenological formula \cite{Yuan, Niesner}.

\begin{equation}
\begin{split}
V_{out}=V_{C}sin(2\varphi)+L_{1}sin(4\varphi)+L_{2}cos(4\varphi)+A
\label{eq:fitting Vcirc}
\end{split}
\end{equation}

Here, $V_{C}$ represents the amplitude associated with the degree of circular polarization of light or helicity (red solid line in Fig. 2), $L_{1}$ is the amplitude associated with the linear polarization of light (blue dash line in Fig. 2), $L_{2}$ depends on the Fresnel coefficients \cite{Okada} (green dash line in Fig. 2) and $A$ is a non-modulated photovoltage offset. $L_{2}$ and $A$ are independent of light polarization. From the fitting we can obtain the amplitude of the photovoltage $V_{out}$, which depends exclusively to the helicity of light, $V_{C}$, and estimate the optical generated spin current by \cite{Mingran, Rojas}

\begin{equation}
\begin{split}
J_{s}=\frac{V_{C}}{\lambda_{IEE}\omega R}
\label{eq:Spin current}
\end{split}
\end{equation}

Where $\lambda_{IEE}$ is the inverse Edelstein effect length directly proportional to the Rashba parameter by $\lambda_{IEE}=(\alpha_{R}\tau_{e})/\hbar$; $\tau_{e}$ is the momentum relaxation time governed by Cu\cite{Tsai}; $\omega$ is the width of our interface and $R$ is the sample resistance. We estimate the spin current by taking the voltage due to circular polarization from fitting of figure 2b, $V_{C}=10.8$$\times10^{-6}$ V, $\lambda_{IEE}=2.35$$\times$10$^{-9}$ m, $\omega=0.9$$\times$10$^{-3}m$ and $R=4.8$$\Omega$ giving a resistivity $\rho=8.64$$\mu$$\Omega cm$, we obtain $J_{s}=1.06$$\times$10$^{6}A/m$$^{2}$, a value comparable with the spin current commonly generated by spin pumping experiments \cite{Mingran, Rojas} and previous reports of circular photovoltaic conversion by inverse spin Hall effect \cite{Isella, Ando2}. This estimation is valid when the contribution from Schottky barrier is negligible. Such is the case for our Cu(111)/$\alpha$-Bi$_{2}$O$_{3}$ \cite{Suppl}. 

In further scrutiny figure 3 compares $V_{C}$ generation in three scenarios: excitation of Cu(111)/$\alpha$-Bi$_{2}$O$_{3}$ by 1.96eV (red dash line), 1.16eV (black dash-dot line) energy lasers and excitation of only Cu layer at 1.96eV (blue solid line), at $\theta=70$$^{\circ}$ and $\Psi=0$$^{\circ}$. Figure 3 shows that $V_{C}$ (1.96eV) $\gg$ $V_{C}$ (1.16eV) for Cu(111)/$\alpha$-Bi$_{2}$O$_{3}$, indicating drastic suppression of the detected photovoltage coming from the circular polarization of light at 1.16eV, and also showing negligible contribution of circular polarized photovoltage coming from Si substrate, which has a band gap of 1.10eV. We also observe that $V_{C}$ (Cu(111)/$\alpha$-Bi$_{2}$O$_{3}$) $\gg$ $V_{C}$ (Cu(111)) with excitation at 1.96eV, indicating the relevance of the Cu(111)/$\alpha$-Bi$_{2}$O$_{3}$ interface. Moreover, we observe an opposite phase of spin to charge conversion for Cu(111)/$\alpha$-Bi$_{2}$O$_{3}$ and Cu(111), in agreement with the opposite sign of spin to charge conversion between Cu(111)/$\alpha$-Bi$_{2}$O$_{3}$ \cite{Tsai, Mingran, Puebla} and the recent reports of conversion of the natural oxide in Cu \cite{An}.  

\begin{figure}[t!]
\begin{center}
\includegraphics[height=6.0cm, keepaspectratio, trim={0 0.3cm 0 0},clip]{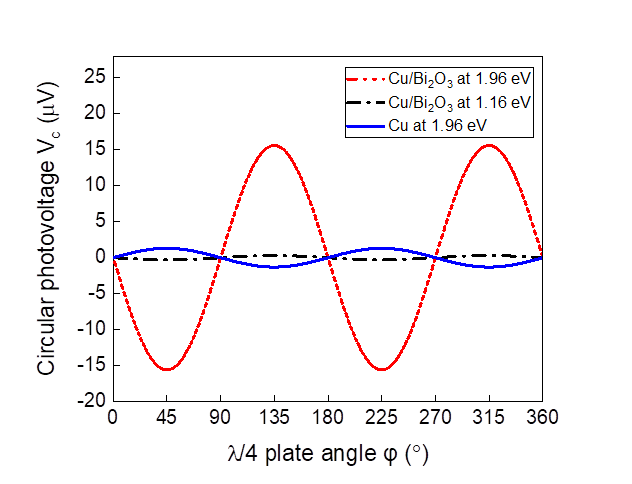}
\end{center}
\caption{Comparison of circularly polarized voltage $V_{C}$. Circular photovoltage conversion of Cu(111)/$\alpha$-Bi$_{2}$O$_{3}$ at 1.96eV (red dash line) is dramatically larger than circular photovoltage conversion at 1.16eV (black dash-dot line), reflecting a threshold energy. Circular photovoltage conversion at 1.96eV of Cu(111)/$\alpha$-Bi$_{2}$O$_{3}$ (red dash line) is significantly larger than circular photovoltage conversion in Cu (blue solid line), reflecting the necessity of an interface with charge transfer and Rashba splitting.}
\label{fig:3}
\end{figure}

First-principle calculations showed hybrization of Cu-O-Bi charge states at our interface and Rashba splitting around 1.98eV above the Fermi energy, allowing the interfacial charge separation mechanism and IEE spin to charge conversion. Transverse photovoltage induced by circular polarized light can be also generated in surface state polaritons via an asymmetric variation of the photon drag effect \cite{Hatano, Bliokh}. This mechanism requires only surface state plasmons in metals and not necessarily the assistance of a semiconductor such as plasmon induced hot electrons mechanism \cite{Tatsuma, Clavero}. We tested the response of a Cu (111) layer to circular polarized light at 1.96eV. While the Cu (111) preserved the optical absorption due to Shockley surface states, we do not observe significant transverse photovoltage related to circular polarized light. Therefore, furthering suggesting combination of interfacial induced charge transfer and IEE as the origin for our circular polarized photovoltage at the Cu(111)/$\alpha$-Bi$_{2}$O$_{3}$ interface. 

\begin{figure}[t!]
\begin{center}
\includegraphics[height=6.0cm, keepaspectratio, trim={0 0.25cm 0 0},clip]{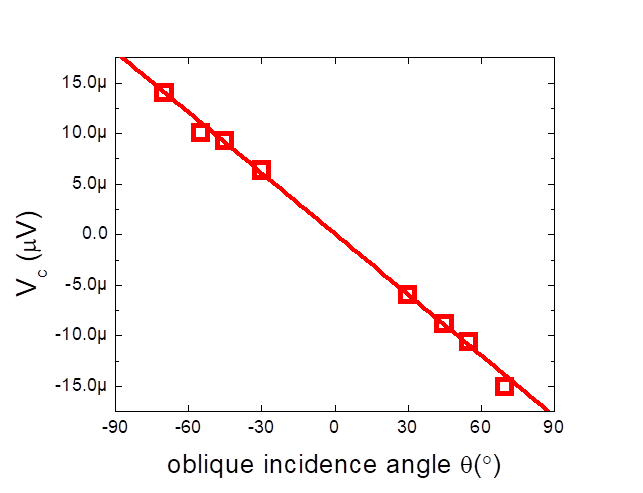}
\end{center}
\caption{Oblique incidence dependence of circularly polarized voltage $V_{C}$. Circular photovoltage conversion of Cu(111)/$\alpha$-Bi$_{2}$O$_{3}$ at 1.96eV (red dash line) increases with the projection onto the interface plane, and changes its sign at opposite oblique incidence angles as expected in spin to charge conversion mechanism.}
\label{fig:4}
\end{figure}

Finally we study the oblique incidence angle dependence of our photoinduced spin to charge conversion (see Fig. 4). The oblique incidence dependence shows an increase of circular photovoltaic signal as the projection is increase onto the plane, and reverses its sign at opposite oblique incidence angles, following spin to charge conversion mechanisms. The interpretation of our data works under the assumption of a interband transition of the $d$-states to the partially filled $s$-states of Cu, as suggested by the typical interband optical absorption, the hybrization of our first principles analysis and the prediction of a Rashba splitting of states at 1.98eV above the Fermi level.

To summarize, we showed the spin photovoltaic conversion at Cu(111)/$\alpha$-Bi$_{2}$O$_{3}$ interface. Due to the increasing number of interfaces with broken spatial inversion symmetry \cite{Ando, Mingran, Tsai, Puebla, Florent, Li}, we expect that the present work motivates further studies, advancing conversion efficiencies and understanding towards spin orbitronics in photovoltaics \cite{Amnu}. From our present and previous reports, we have indication of spin to charge conversion at Cu(111)/$\alpha$-Bi$_{2}$O$_{3}$ interface due to Rashba spin orbit coupling \cite{Mingran, Tsai, Puebla, Florent, Rojas}. Rashba spin orbit coupling is suggested as key component to suppress carrier recombination and enhanced carrier lifetime in perovskites \cite{Even, Zheng}. We observed an efficient photovoltaic conversion arising from a charge transfer mechanism at our Cu(111)/$\alpha$-Bi$_{2}$O$_{3}$ interface and Rashba spin splitting in an excited state. The photoinduced spin to charge conversion via Rashba spin splitting in an excited state motivates further studies in similar structures, and further understanding of the mechanism involved. Very recently, a related report shows circular photovoltaic signal at metal/metal interface \cite{Hirose}, the interpretation of our results may shed new light in the understanding of this recent report and motive further studies. Our device is compatible with complementary metal-oxide-semiconductor (CMOS) technology, opening a new venue for exploring spin orbitronics at interfaces towards spin electronic devices beyond the Moore’s law \cite{Intel}. 

\hfill \break
We acknowledge Naoki Ogawa and Kouta Kondo for fruitful discussions. This work was supported by Grant-in-Aid for Scientific Research on Innovative Area, Nano Spin Conversion Science (Grant No. 26103002 and No. 17H05180) and RIKEN Incentive Research Project Grant No. FY2016. The first-principles calculation was supported in part by MEXT as a social and scientific priority issue (Creation of new functional devices and high-performance materials to support next-generation industries) to be tackled by using post-K computer (Project ID: hp180206). F.A. was supported by the Ministry of Education, Culture, Sports, Science and Technology (MEXT) Scholarship, Japan

%\addcontentsline{toc}{chapter}{Bibliography}

\hfill \break

\section{Appedix}

\subsection{Methods}

\subsubsection{Device fabrication and experimental setup}

Our samples consist of a Cu(30nm)/Bi$_{2}$O$_{3}$(20nm) bilayer prepared on a Si substrate with a 300nm thick SiO$_{2}$ layer. The sample is patterned by using mask-less UV-Lithography. The resists consist of a primer (hexamethyl-disilazane) and AZ1500, both are coated at 4000 rpm for 45 seconds and baked respectively for 5min and 10min. Depositions are done in-situ using e-beam evaporation at a vacuum pressure of 5$\times$10$^{-5}$ Pa at a rate of 1 \AA/s for Cu and 0.2 \AA/s for Bi$_{2}$O$_{3}$. Two electrodes made of Au(150nm)/Ti(5nm), also deposited by e-beam evaporation, are added to measure the voltage generated by illuminating a laser beam onto the sample. The laser beam has an incidence angle $\theta$ and an azimuthal angle $\Psi$.  Our light excitations are continuous wave lasers at energies of 1.16eV and 1.96 eV. The laser spots are adjusted to approximately 400$\mu$m for all the lasers, and placed at the center of our sample surface to minimize the effect of thermal gradients. The photon polarization is controlled by a linear polarizer and a quarter wave plate mounted on a mechanical rotator. The voltage is detected by a lock-in amplifier in open circuit mode synced at 400 Hz with a mechanical chopper.

\subsubsection{First-principles calculation (Computational details)}

Our density functional calculations were performed using OpenMX code \cite{Ozaki, Ozaki2} within the generalized gradient approximation \cite{Perdew}. We used 16$\times$12$\times$1 regular k-point mesh and the fully relativistic total angular momentum dependent pseudo-potentials taking spin-orbit interaction (SOI) into account \cite{Theu}. We adopted norm-conserving pseudopotentials with an energy cutoff of 300 Ry for charge density including the 5d, 6s and 6p-states as valence states for Bi; 2s and 2p for O; 3s, 3p, 3d and 4s for Cu.

The numerical pseudo atomic orbitals \cite{Ozaki3} are used as follows: the numbers of the s-, p- and d-character orbitals are three, three and two, respectively; The cutoff radii of Bi, O, and Cu are 8.0, 5.0 and 6.0, respectively, in units of Bohr. The dipole-dipole interaction between slab models can be eliminated by the effective screening medium (ESM) method \cite{OtaniM, Oh}. A Rashba energy E$_{R}$ and a Rashba momentum k$_{R}$ were obtained from a difference in energy and a distance in momentum space, respectively between the band crossing point, that is, C point (0.5, 0.5) and the band bottom point.

\subsection{Additional data fittings}

\subsubsection{Polarization dependent photovoltage of Cu and Cu/Bi$_{2}$O$_{3}$}

In figure 3 of our main text we showed a comparison of circular polarization photovoltage for Cu/Bi$_{2}$O$_{3}$ excited by laser energy of 1.96eV, 1.16eV and Cu layer excited by laser energy of 1.96eV. The circular polarization photovoltage is extracted from fitting following eq. (1) of our main text. For completeness, here we provide full fitting plots of similar device under same conditions (figure 5). 

For our present study the most relevant component is the photovoltage associated to the circular polarization of light (V$_{C}$). Similar to our results in the main text, we find that the maximum circular polarization photovoltage occurs for Cu/Bi$_{2}$O$_{3}$ excited by laser energy of 1.96eV. We also observed that the sign for circular polarization photovoltage for Cu/Bi$_{2}$O$_{3}$ and Cu layer have opposite signs, suggesting opposite spin to charge photo-conversion, in agreement with previous reports of spin to charge conversion in natural oxide in Cu \cite{An}. 

\begin{figure}
\begin{center}
\includegraphics[height=6.0cm, keepaspectratio,  trim={0 2.0cm 0 0}, clip]{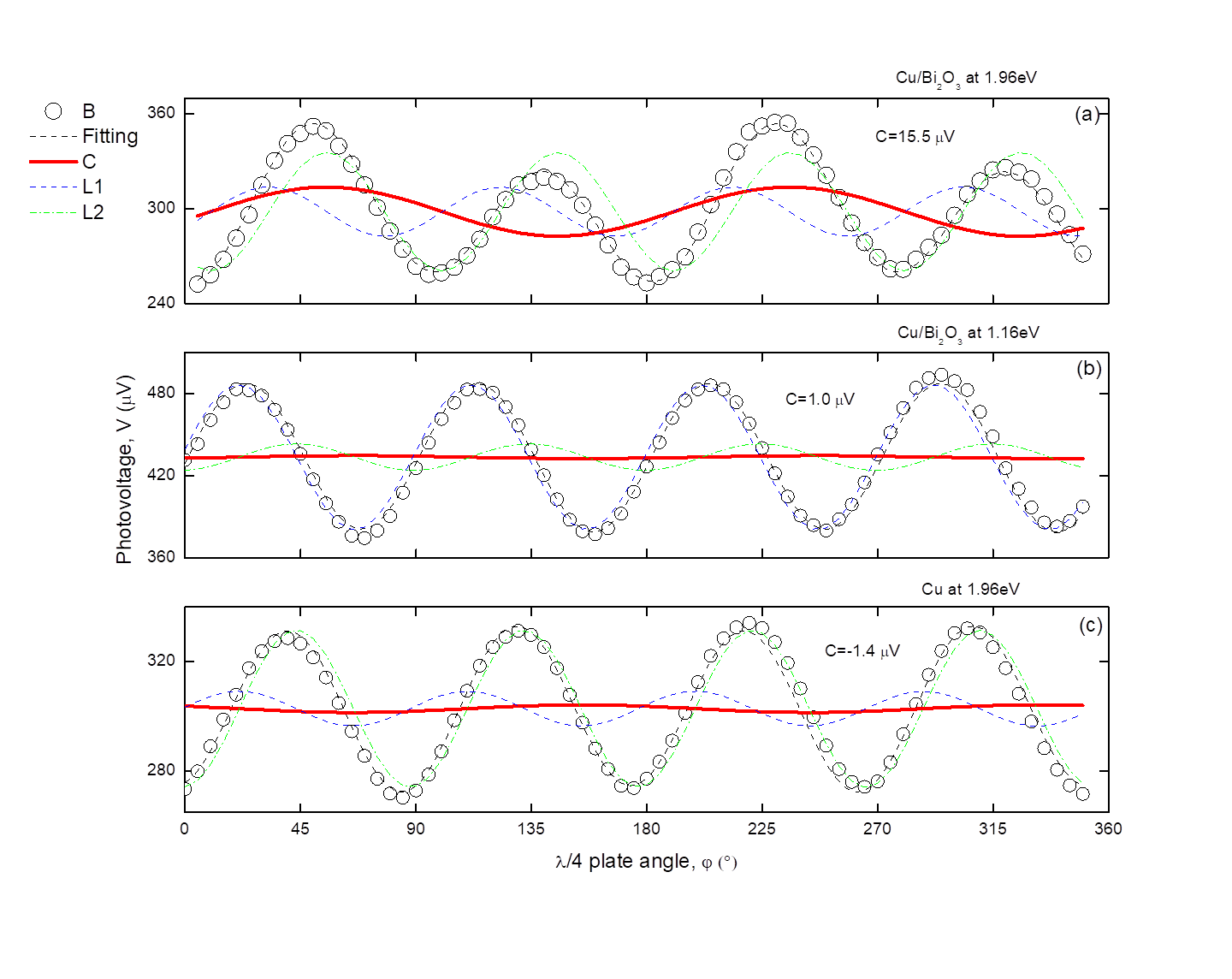}
\end{center}
\caption{Helicity dependent photovoltage. Polarization dependent photovoltage for Cu/Bi$_{2}$O$_{3}$ excited by laser energy of 1.96eV (a), 1.16eV (b) and Cu (c) layer excited by laser energy of 1.96eV. Fitting follows eq. (1) of our main text.}
\label{fig:4}
\end{figure}

Additional to the circular polarization photovoltage V$_{C}$, we also extract the two linear polarization photovoltage, L1 and L2. These two linear polarization components are spin independent, therefore out of the main scope of our present study. Nevertheless, we find appropriate to make some brief comments of our linear polarization photovoltages. In a general form, both the circular photovoltaic effect (V$_{C}$ in our manuscript) and linear photovoltaic effect (L1 and L2 in our manuscript) exist in non-centrosymmetric systems, meaning systems with spatial inversion symmetry breaking. While the circular photovoltaic effect is time reversal invariant, the linear photovoltaic effect is time reversal variant. Under homogeneous excitation and transverse geometry of incidence (the geometry of our experiments), there are two main sources of linear photovoltage: 1) the scattering of free carriers in phonons, defects and other carriers in noncentrosymmetric media (L1); and 2) The optical response and momentum transfer of s and p linear polarization, related to the Fresnel coefficients, as indicated in our manuscript (L2).  Since the linear polarization is significantly affected by the refractive index of materials and the interaction of light with scattering centers, the presence or absence of a top Bi$_{2}$O$_{3}$ layer and changes of the excitation energy may significantly modify the linear polarization components L1 and L2, as we indeed observe in figure 5.   
For completeness, in figure 6 we provide an additional oblique incidence dependence plot of circular photovoltage excited at 1.96eV of a Cu/Bi$_{2}$O$_{3}$ device from a different batch to the data of figure 4 of our main text. 

\begin{figure}
\begin{center}
\includegraphics[height=6.0cm, keepaspectratio,  trim={0 1.0cm 0 0}, clip]{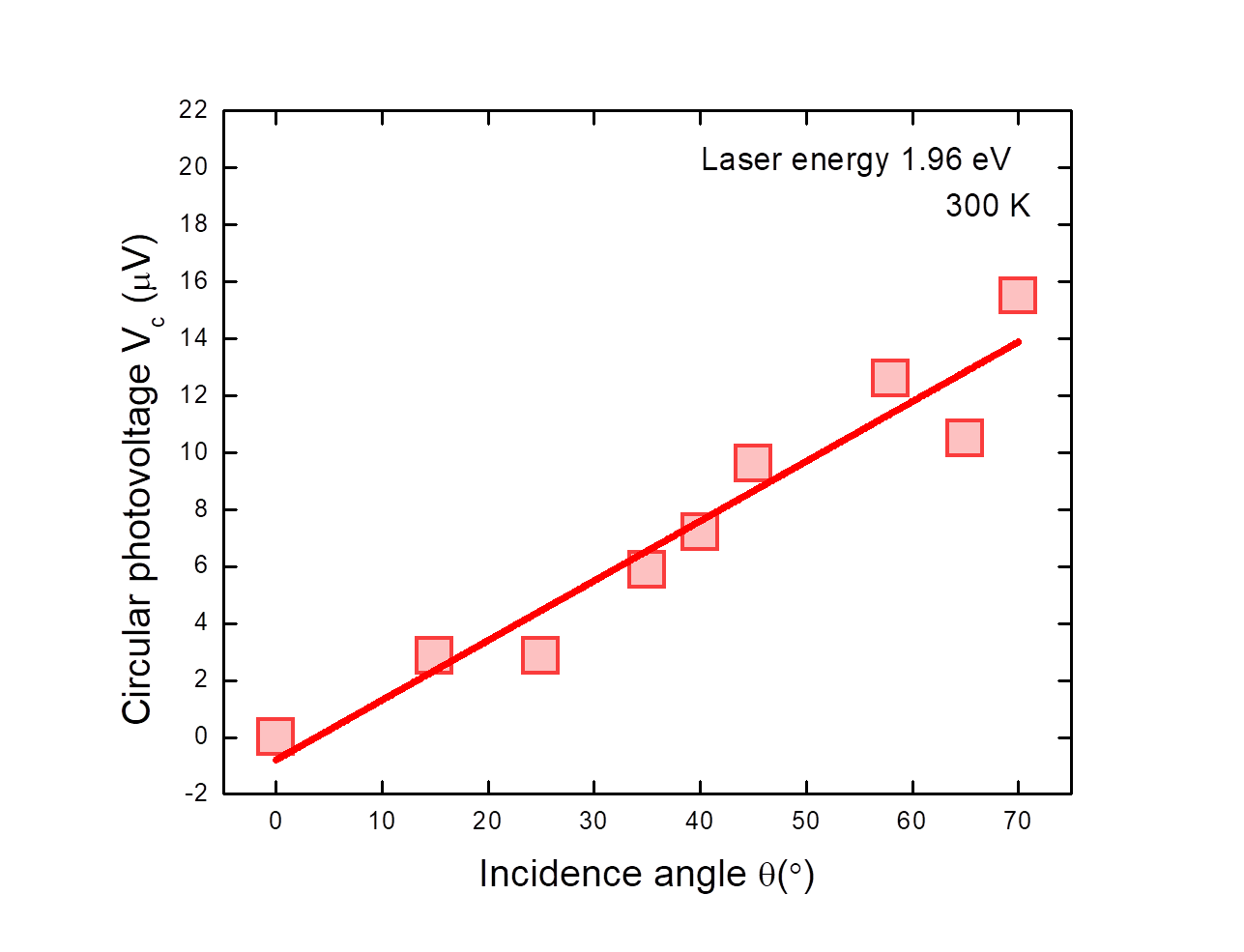}
\end{center}
\caption{Oblique incidence dependence of circularly polarized voltage, V$_{C}$. Circular photovoltage conversion of Cu(111)/Bi$_{2}$O$_{3}$ at 1.96eV increases with the projection onto the interface plane.}
\label{fig:4}
\end{figure}

\subsection{I-V characterization}

We measured I-V characteristics of our Cu/Bi$_{2}$O$_{3}$ heterojunction. I-V measurement indicates the formation of an Ohmic heterojunction, which minimize the influence of Schottky barrier rectification as mechanism of our photovoltaic conversion. I-V measurements were done using a stacking of top contact Ti(5nm)/Au(150nm), Cu(30nm)/Bi$_{2}$O$_{3}$(20nm), bottom contact Au(300nm), 200nm of Al$_{2}$O$_{3}$ were also deposited to avoid leakage current. Figure 7 shows I-V measurement with Ohmic junction behavior between Cu and Bi$_{2}$O$_{3}$ for samples with different contact area.

\begin{figure}
\begin{center}
\includegraphics[height=6.0cm, keepaspectratio,  trim={0 1.25cm 0 0}, clip]{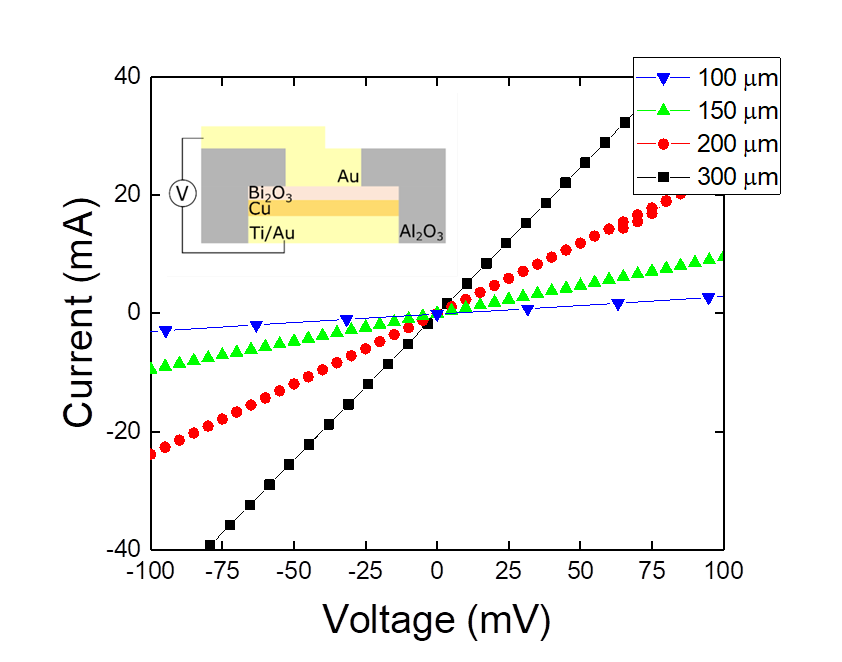}
\end{center}
\caption{I-V characteristics. I-V characteristics of the Cu/Bi$_{2}$O$_{3}$ heterojunction for different contact areas. Inset shows schematic of measurement.}
\label{fig:4}
\end{figure}

\subsection{Optical absorption spectroscopy}

Optical absorption of Cu/Bi$_{2}$O$_{3}$ device by UV-Vis-NIR absorption spectroscopy (Shimadzu UV-3600 Plus) in the range from 0.9eV to 4eV, see figure 8. Figure 8a shows the Tauc plot of Bi$_{2}$O$_{3}$ (20nm) grown on sapphire (0001) showing a band gap at 3.21eV. Figure 8b shows the Transmission, Reflection and Absorption of Cu (30nm). Fig 8c shows the thickness dependence of Cu(X)/Bi$_{2}$O$_{3}$. The drastic change of absorption around 2eV corresponds to the interband transition from the 3d bands to the 4s band of Cu. Figure 8d shows the Tauc plot of Cu(30nm)/Bi$_{2}$O$_{3}$ showing a shift of band gap energy of Bi$_{2}$O$_{3}$ at 3.09 eV.   

\begin{figure}
\begin{center}
\includegraphics[height=6.0cm, keepaspectratio,  trim={0 1.0cm 0 0}, clip]{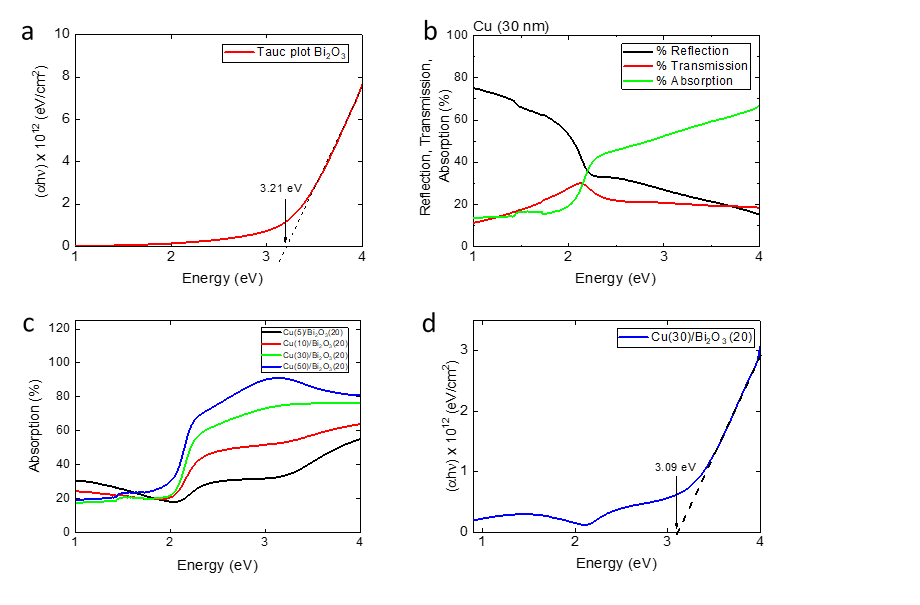}
\end{center}
\caption{Optical absorption. UV-Vis-NIR absorption spectroscopy in the range from 0.9eV to 4eV of our Cu/Bi$_{2}$O$_{3}$. a. Tauc plot of Bi$_{2}$O$_{3}$ (20 nm) grown on sapphire (0001) showing a band gap of Bi$_{2}$O$_{3}$ at 3.21 eV. b. Transmission, Reflection and Absorption of Cu (30nm). c. Thickness dependence of Cu(X)/Bi$_{2}$O$_{3}$. d. Tauc plot of Cu/Bi$_{2}$O$_{3}$ showing a shift of band gap of Bi$_{2}$O$_{3}$ at 3.09 eV.}
\label{fig:4}
\end{figure}

\subsection{XRD spectroscopy}

\begin{figure}[H]
\begin{center}
\includegraphics[height=6.0cm, keepaspectratio,  trim={0 0.5cm 0 0.5cm}, clip]{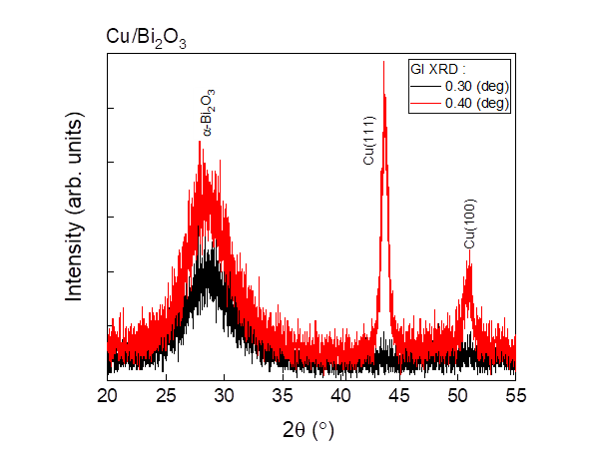}
\end{center}
\caption{XRD spectroscopy. X-ray diffraction spectroscopy measurements for Cu/Bi$_{2}$O$_{3}$ at grazing incident (GI). Broad peak at $\theta=$28$^{\circ}$ reflects the $\alpha$-phase of Bi$_{2}$O$_{3}$ layer, the Cu layer at the interface show preferential orientation at the interface in the (111) direction.}
\label{fig:4}
\end{figure}

X-ray diffraction at grazing incident (GI) spectroscopy measurements were done for Cu/Bi$_{2}$O$_{3}$ with Rigaku SmartLab high resolution X-ray diffractometer (Fig.9). At an angle of 0.30$^{\circ}$, GI incidence (black line), only one feature appears, a broad peak at $\theta=$28$^{\circ}$ at the interface of Cu/Bi$_{2}$O$_{3}$. This broad peak is the characteristic peak of our top Bi$_{2}$O$_{3}$ layer, and it indicates a amorphous structure in $\alpha$-phase, the most stable phase at room temperature for Bi$_{2}$O$_{3}$. At a GI angle of 0.40$^{\circ}$ (red line), the Cu layer shows preferential crystallinity at the interface in the (111) direction. 

\end{document}